\documentclass[pra,aps,twocolumn,superscriptaddress,showpacs,showkeys,nofootinbib]{revtex4-1}
\usepackage[latin1]{inputenc}
\usepackage{amsmath,amsthm,amssymb,graphicx,subfigure,xcolor}
\allowdisplaybreaks[4]
\usepackage{mathrsfs}
\usepackage{graphics,float}
\usepackage{epstopdf}
\usepackage{color}
\usepackage[unicode=true]{hyperref}
\usepackage{booktabs}
\usepackage{tabularx}
\usepackage{tabu}
\usepackage{multirow}
\hypersetup{
     colorlinks=true,       		
     linkcolor=blue,          	
     citecolor=blue,            
     urlcolor=blue,           	
 }

\newtheorem*{theorem*}{Theorem}

\newtheorem*{corollary*}{Corollary}

\newtheorem*{lemma*}{Lemma}

\newtheorem*{proposition*}{Proposition}
\theoremstyle{definition}

\newtheorem*{definition*}{Definition}
\theoremstyle{remark}

\newtheorem*{remark*}{Remark}

\begin{document}
\title{$\boldsymbol{G_q}$-concurrence and entanglement constraints in multiqubit systems}
\author{Hui Li}
\author{Ting Gao}
\email{gaoting@hebtu.edu.cn}
\affiliation{School of Mathematical Sciences, Hebei Normal University, Shijiazhuang 050024, China}
\author{Fengli Yan}
\email{flyan@hebtu.edu.cn}
\affiliation{College of Physics, Hebei Key Laboratory of Photophysics Research and Application, Hebei Normal University, Shijiazhuang 050024, China}
\begin{abstract}
In this paper, we introduce a category of one-parameter bipartite entanglement quantifiers, termed $G_q$-concurrence ($q>1$), and show rigorously that they satisfy all the axiomatic conditions of an entanglement measure and can be considered as a generalization of concurrence. In addition, we establish an analytic formula relating $G_q$-concurrence to concurrence for $1<q\leq2$ in two-qubit systems. Furthermore, the polygamy relation is presented based on the $G_q$-concurrence of assistance in multiqubit systems. As far as $G_q$-concurrence ($1<q\leq2$) itself is concerned, however, it does not obey the monogamy relation, but we prove that the square of $G_q$-concurrence does. By means of this monogamy inequality, we construct a set of entanglement indicators that can detect genuinely multiqubit entangled states even when the tangle loses its efficacy.
\end{abstract}



\maketitle

\section{Introduction}\label{I}
Quantum entanglement, as an essential physical resource, exhibits prominent advantages over classical theory in quantum information tasks, such as quantum computation \cite{45,46,47}, quantum sensing \cite{48}, and quantum communication \cite{18,19,20,21,22,23}. A fundamental and necessary task in entanglement resource theory is undoubtedly the development of a legitimate method for quantifying the entanglement of states. In addition, another imperative assignment is to explore the intrinsic properties of a measure, such as whether it satisfies monogamy of entanglement (MoE).

MoE means that there are some restrictions on shareability and distribution of entanglement \cite{6,37}. The mathematical characterization of monogamy relation was first introduced by Coffman $et~al.$ \cite{11} based on the square of concurrence \cite{9,7} in three-qubit systems. Subsequently, Osborne and Verstraete \cite{2} extended it to multiqubit systems. Furthermore, various efforts have been dedicated to the study of MoE \cite{33,30,34,31,36,35,24,28,29}, and analogical monogamy inequalities have been rendered in terms of Tsallis-$q$ entanglement \cite{8,32}, entanglement of formation \cite{13,26}, and R\'{e}nyi-$\alpha$ entanglement \cite{25,27}.

Entanglement of assistance, a dual notion of bipartite entanglement measure, indicates how much entanglement two parties can share through the assistance of another party or other parties \cite{38,39}. Gour $et~al$. proved that concurrence of assistance (CoA) obeys polygamy relation for three-qubit pure states \cite{39}, and then it was generalized to multiqubit scenario \cite{38}.

Although there are a battery of entanglement measures to describe the entanglement distribution of multipartite quantum states \cite{43,4,12,40,44}, the difficulty of calculating how much entanglement a mixed state contains remains. However, MoE, in fact, offers an upper bound for bipartite sharability of entanglement in multipartite systems, whereas its dual (polygamy of entanglement) gives a lower bound for distribution of bipartite entanglement. Moreover, MoE also has momentous applications in other areas of physics, such as quantum cryptography \cite{14,15,16,17}, condensed matter physics \cite{41}, and quantum channel discrimination \cite{42}.

It is admitted that tangle generated from MoE of the concurrence squared fails to detect the entanglement of some states, such as the $W$ state \cite{11}. To compensate this deficiency, other novel monogamy relations beyond concurrence squared should be established. Therefore, we construct a class of bipartite entanglement measures $G_q$-concurrence ($q>1$) and verify their squares conform to monogamy inequality with $1<q<2$. Most importantly, the series of indicators produced by MoE of the $G_q$-concurrence can identify all genuinely multiqubit entangled states.

The paper is structured as follows. In Sec.\ref{II}, we define $G_q$-concurrence and its dual quantity, where $G_q$-concurrence complies with the requirements including faithfulness, invariance under any local unitary transformation, (strong) monotonicity, convexity, and subadditivity. In particular, concurrence can be perceived as a special form of $G_q$-concurrence corresponding to $q=2$. An analytic relation between $G_q$-concurrence and concurrence, in Sec.\ref{III}, is provided for $1<q\leq2$ in two-qubit quantum systems. In Sec.\ref{IV}, we derive that $G_q$-concurrence of assistance follows polygamy inequality for $1<q\leq2$ in multiqubit systems. In Sec.\ref{V}, we show that the square of $G_q$-concurrence obeys monogamy relation and construct a set of entanglement indicators. Besides, the monogamy property of the $\alpha$-th ($\alpha\geq2$) power of $G_q$-concurrence is discussed. The main conclusions are summarized in Sec.\ref{VI}.

\section{$\boldsymbol{G_q}$-concurrence}\label{II}
In this section, we first define a kind of  bipartite entanglement quantifiers, called $G_q$-concurrence, which can be counted as a generalization of concurrence. Moreover, we prove that $G_q$-concurrence possesses several fundamental properties of an entanglement measure.

{\bf Definition 1}. For any bipartite pure state $|\phi\rangle_{AB}$, the $G_q$-concurrence is defined as
\begin{equation}\label{1}
\begin{array}{rl}
\mathscr{C}_q(|\phi\rangle_{AB})=[1-{\rm Tr}(\rho_{A}^q)]^{\frac{1}{q}}
\end{array}
\end{equation}
for $q>1$. Here $\rho_{A}={\rm Tr}_B(|\phi\rangle_{AB}\langle\phi|)$.

A bipartite pure state $|\phi\rangle_{AB}$ can be expressed in the Schmidt decomposition form
\begin{equation}\label{8}
\begin{array}{rl}
|\phi\rangle_{AB}=\sum_{i=1}^d\sqrt{\lambda_i}|i_A\rangle|i_B\rangle,
\end{array}
\end{equation}
where $\lambda_i$ is non-negative real number with $\sum_i\lambda_i=1$. Then, $\mathscr{C}_q(|\phi\rangle_{AB})$ can be written as
\begin{equation}\label{9}
\begin{array}{rl}
\mathscr{C}_q(|\phi\rangle_{AB})=(1-\sum_{i=1}^{d}\lambda_i^q)^{\frac{1}{q}},~q>1.
\end{array}
\end{equation}

For any bipartite mixed state $\rho_{AB}$, its $G_q$-concurrence is defined as
\begin{equation}\label{2}
\begin{array}{rl}
\mathscr{C}_{q}(\rho_{AB})=\min\limits_{\{p_i,|\phi_{i}\rangle\}}\sum\limits_{i}p_i\mathscr{C}_{q}(|\phi_i\rangle),
\end{array}
\end{equation}
where the minimum is taken over all possible ensemble decompositions $\{p_i,|\phi_{i}\rangle\}$ of $\rho_{AB}$.

In particular, $G_q$-concurrence $\mathscr{C}_q(\rho_{AB})$ and concurrence $C(\rho_{AB})$ are equivalent when $q=2$ for any bipartite quantum state $\rho_{AB}$. Therefore, $G_q$-concurrence can be treated as a generalization of concurrence \cite{1,6,7,9}. Moreover, the relation between $G_q$-concurrence and $q$-concurrence \cite{10} is $\mathscr{C}_q(|\phi\rangle_{AB})=[C_q(|\phi\rangle_{AB})]^{1/q}$ for any bipartite pure state $|\phi\rangle_{AB}$.

The $G_q$-concurrence of assistance ($G_q$-CoA), a dual quantity of $G_q$-concurrence, is given by
\begin{equation}\label{23}
\begin{array}{rl}
\mathscr{C}_{q}^a(\rho_{AB})=\max\limits_{\{p_i,|\phi_{i}\rangle\}}\sum\limits_{i}p_i\mathscr{C}_{q}(|\phi_i\rangle),
\end{array}
\end{equation}
where the maximum runs over all feasible ensemble decompositions of $\rho_{AB}$. If $\rho_{AB}$ is a pure state, then there is $\mathscr{C}_{q}^a(\rho_{AB})=\mathscr{C}_{q}(\rho_{AB})$.

Some studies suggest that a rational entanglement measure should satisfy several conditions \cite{6}, including:

(i) Faithfulness;

(ii) Invariance under any local unitary transformation;

(iii) Monotonicity under local operation and classical communication (LOCC).

We will elaborate in the following subsections that $G_q$-concurrence not only satisfies the conditions (i) to (iii) listed above, but also fulfills the properties as follows,

(iv) Entanglement monotone \cite{3} (or strong monotonicity under LOCC);

(v) Convexity;

(vi) Subadditivity.

\subsection{Faithfulness}
Faithfulness is an essential property for entanglement quantifiers, which can clearly distinguish bipartite quantum states into two categories, entangled states and separable states. Next we demonstrate that $G_q$-concurrence is faithful.

{\bf Proposition 1}. For any bipartite quantum state $\rho_{AB}$, we have $\mathscr{C}_q(\rho_{AB})\geq0$ for $q>1$, the equality holds iff $\rho_{AB}$ is a separable state.

{\bf Proof}. It is obvious that $\mathscr{C}_q(\rho_{AB})\geq0$ since ${\rm Tr}(\rho_A^q)\leq1$ for $q>1$.

Next, we first prove the equality is true iff $|\phi\rangle_{AB}$ is a separable state. If a pure state $|\phi\rangle_{AB}$ is separable, then we can get ${\rm Tr}(\rho_A^q)=1$, which leads $\mathscr{C}_q(|\phi\rangle_{AB})=0$. Conversely, let $|\phi\rangle_{AB}=\sum_i\sqrt{\lambda_i}|i_A\rangle|i_B\rangle$, one has the reduced density operator $\rho_A=\sum_i\lambda_i|i_A\rangle\langle i_A|$. If $\mathscr{C}_q(|\phi\rangle_{AB})=0$, then the Schmidt number of $|\phi\rangle_{AB}$ must be one due to $0\leq\lambda_i\leq1$ and $q>1$, i.e., $|\phi\rangle_{AB}=|i_A\rangle|i_B\rangle$, hence the pure state $|\phi\rangle_{AB}$ is separable.

For any separable mixed state $\rho_{AB}$ with the pure decomposition $\{p_i,|\phi_i\rangle_{AB}\}$, $\mathscr{C}_q(\rho_{AB})\leq\sum_ip_i\mathscr{C}_q(|\phi_i\rangle_{AB})=0$, owing to the nonnegativity of $\mathscr{C}_q(\rho_{AB})$, we have $\mathscr{C}_q(\rho_{AB})=0$. On the contrary, if $\mathscr{C}_q(\rho_{AB})=0$, according to definition of $G_q$-concurrence, one has $\mathscr{C}_q(|\phi_i\rangle_{AB})=0$ for any $i$, which is equivalent to $|\phi_i\rangle_{AB}$ being separable for every $i$, so $\rho_{AB}$ is separable.

To sum up, $\mathscr{C}_q(\rho_{AB})>0$ for all entangled states and $\mathscr{C}_q(\rho_{AB})=0$ for all separable states. $\hfill\blacksquare$

\subsection{Monotonicity under LOCC and invariance under local unitary transformation}
In Ref. \cite{3}, Vidal put forward the only necessary condition on the entanglement measure should be that entanglement does not increase under LOCC. We will verify that $G_q$-concurrence obeys this requirement.

{\bf Proposition 2}. Let $\rho_{AB}$ be any bipartite state, the $G_q$-concurrence is non-increasing under LOCC operation $\Lambda_{\rm LOCC}$, i.e., $\mathscr{C}_q[\Lambda_{\rm LOCC}(\rho_{AB})]\leq \mathscr{C}_q(\rho_{AB})$.

{\bf Proof}. The state $|\phi\rangle$ can be prepared from the state $|\varphi\rangle$ using only LOCC iff the vector $\vec{\lambda}_\phi$ majorizes $\vec{\lambda}_\varphi$ ($\vec{\lambda}_\varphi\prec\vec{\lambda}_\phi$), where $\vec{\lambda}_\varphi$ ($\vec{\lambda}_\phi$) is Schmidt vector given by the squared Schmidt coefficients of the state $|\varphi\rangle$ ($|\phi\rangle$) and arrange in non-increasing order \cite{1}. A function $E$ is monotone on pure state iff it is Schur concave as a function of spectrum of subsystem, which is equivalent to the following two conditions \cite{5}: (a) $E$ is invariant under any permutation of two arguments; (b) any two components of $\vec{\lambda}$, $\lambda_i$ and $\lambda_j$, satisfy $(\lambda_i-\lambda_j)(\frac{\partial E}{\partial\lambda_i}-\frac{\partial E}{\partial\lambda_j})\leq0$.

We first show that the function $\mathscr{C}_q(|\phi\rangle)$ is non-increasing under LOCC with respect to any bipartite pure state $|\phi\rangle$ using the methods described above. Let $\mathscr{C}_q(|\phi\rangle)=f(\lambda_1,\lambda_2,\cdots,\lambda_d)=(1-\sum_{i=1}^{d}\lambda_i^q)^{\frac{1}{q}}$, where $\lambda_1,\lambda_2,\cdots,\lambda_d$ are the square of Schmidt coefficients of $|\phi\rangle$ and satisfy $\lambda_1\geq\lambda_2\geq\cdots\geq\lambda_d$. It is easy to get that $\mathscr{C}_q(|\phi\rangle)$ is invariant when any two arguments $\lambda_i$ and $\lambda_j$ of the vector $\vec{\lambda}_\phi$ permute, and
\begin{align*}
&(\lambda_i-\lambda_j)(\frac{\partial \mathscr{C}_q}{\partial\lambda_i}-\frac{\partial \mathscr{C}_q}{\partial\lambda_j})\\
=&(\lambda_i-\lambda_j)[(1-\sum\limits_{l=1}^{m}\lambda_l^q)^{\frac{1}{q}-1}\lambda_j^{q-1}-(1-\sum\limits_{l=1}^{m}\lambda_l^q)^{\frac{1}{q}-1}\lambda_i^{q-1}]\\
=&(\lambda_i-\lambda_j)(1-\sum\limits_{l=1}^{m}\lambda_l^q)^{\frac{1}{q}-1}(\lambda_j^{q-1}-\lambda_i^{q-1})\\
\leq&0.
\end{align*}
Thus, we conclude that $\mathscr{C}_q[\Lambda_{\rm LOCC}(|\phi\rangle)]\leq\mathscr{C}_q(|\phi\rangle)$.

Adopting the convexity of $\mathscr{C}_{q}(\rho_{AB})$ and the monotonicity of $\mathscr{C}_{q}(|\phi\rangle)$, we can obtain the result that $G_q$-concurrence does not increase under LOCC for any bipartite mixed state. $\hfill\blacksquare$

It is known that local unitary transformations belong to the set of LOCC operations and are invertible \cite{6}. From Proposition 2, we proceed directly to the following conclusion.

{\bf Proposition 3}. For any bipartite quantum state $\rho_{AB}$, $\mathscr{C}_q(\rho_{AB})$ is invariant under any local unitary transformation, i.e., $\mathscr{C}_q(\rho_{AB})=\mathscr{C}_q(U_A\otimes U_B\rho_{AB}U_A^\dagger\otimes U_B^\dagger)$.

\subsection{Entanglement monotone}
Before proving the strong monotonicity of $G_q$-concurrence, let us present a lemma.

{\bf Lemma 1}. The function
\begin{equation}
\begin{array}{rl}
{G}_q(\rho)=(1-{\rm Tr}\rho^q)^{\frac{1}{q}}
\end{array}
\end{equation}
is concavity for any density operator $\rho$ and $q>1$.

{\bf Proof}. Let $\rho,\sigma$ be two arbitrary density operators, we derive
\begin{equation}
\begin{array}{rl}
{G}_q(\lambda\rho+\mu\sigma)
=&[1-{\rm Tr}(\lambda\rho+\mu\sigma)^q]^{\frac{1}{q}}\\
\geq&\{1-[\lambda({\rm Tr}\rho^q)^{\frac{1}{q}}+\mu({\rm Tr}\sigma^q)^{\frac{1}{q}}]^q\}^{\frac{1}{q}}\\
\geq&[1-(\lambda{\rm Tr}\rho^q+\mu{\rm Tr}\sigma^q)]^{\frac{1}{q}}\\
=&[\lambda(1-{\rm Tr}\rho^q)+\mu(1-{\rm Tr}\sigma^q)]^{\frac{1}{q}}\\
\geq&\lambda(1-{\rm Tr}\rho^q)^{\frac{1}{q}}+\mu(1-{\rm Tr}\sigma^q)^{\frac{1}{q}}\\
=&\lambda{G}_q(\rho)+\mu{G}_q(\sigma),\\
\end{array}
\end{equation}
where the first inequality can be obtained based on Minkowski's inequality $[{\rm Tr}(\rho+\sigma)^q]^{\frac{1}{q}}\leq({\rm Tr}\rho^q)^{\frac{1}{q}}+({\rm Tr}\sigma^q)^{\frac{1}{q}}$ with $q>1$, the second inequality holds because the function $y=x^q$ is convex for $q>1$, and the third inequality is due to the concavity of $y=x^\gamma$ for $0<\gamma<1$. $\hfill\blacksquare$

{\bf Proposition 4}. For any bipartite state $\rho_{AB}$, the $G_q$-concurrence is an entanglement monotone, namely,
\begin{equation}\label{3}
\begin{array}{rl}
\mathscr{C}_q(\rho_{AB})\geq\sum_ip_i\mathscr{C}_q(\sigma_i),\\
\end{array}
\end{equation}
where the ensemble $\{p_i,\sigma_i\}$ is yielded after $\Lambda_{\rm LOCC}$ acting on $\rho_{AB}$.

{\bf Proof}. Vidal \cite{3} showed that an entanglement quantifier $E$ obeys strong monotonicity if it satisfies the following two conditions: (c) $g(U\rho_AU^\dagger)=g(\rho_A)$ and $g$ is a concave function, where $E(|\phi\rangle_{AB})=g(\rho_A)$ and $\rho_A={\rm Tr}_B(|\phi\rangle\langle\phi|)$; (d) $E$ is given by convex roof extension for arbitrary mixed states. It is obvious that $\mathscr{C}_{q}(\rho_{AB})$ meets these conditions from Proposition 3, Lemma 1, and the definition of $\mathscr{C}_{q}(\rho_{AB})$. Therefore, the formula (\ref{3}) holds. $\hfill\blacksquare$

\subsection{Convexity}
According to the definition of $\mathscr{C}_q(\rho_{AB})$, the following result can be reached.

{\bf Proposition 5}. The $G_q$-concurrence is convex on quantum state $\rho_{AB}$, that is, $\mathscr{C}_q(\rho_{AB})\leq\sum_ip_i\mathscr{C}_q(\rho^i_{AB})$, where $\rho_{AB}=\sum_ip_i\rho^i_{AB}$, $\sum_ip_i=1$, and $p_i>0$.

\subsection{Subadditivity}
{\bf Proposition 6}. The $G_q$-concurrence is subadditive, i.e., $\mathscr{C}_q(\rho_{AB}\otimes\sigma_{AB})\leq \mathscr{C}_q(\rho_{AB})+\mathscr{C}_q(\sigma_{AB})$.

{\bf Proof}. Before proving subadditivity, we first show the inequality $(a+b)^\beta\leq a^\beta+b^\beta$ holds for $0\leq a,b\leq1$ and $0<\beta<1$, which is equivalent to
\begin{equation}\label{4}
\begin{array}{rl}
(\frac{a}{a+b})^\beta+(\frac{b}{a+b})^\beta\geq1.\\
\end{array}
\end{equation}
When $0\leq x_1,x_2\leq1$ and $x_1+x_2=1$, we have $x_1^\beta\geq x_1$ and $x_2^\beta\geq x_2$, and this goes directly to inequality $x_1^\beta+x_2^\beta\geq1$. Let $x_1=\frac{a}{a+b}$ and $x_2=\frac{b}{a+b}$, the formula (\ref{4}) can be obtained.

For any two pure states $|\phi\rangle_{AB}$ and $|\varphi\rangle_{AB}$, then we can see
\begin{equation}\label{5}
\begin{array}{rl}
&\mathscr{C}_q(|\phi\rangle_{AB}\otimes|\varphi\rangle_{AB})\\
=&(1-{\rm Tr}\rho_{A}^q{\rm Tr}\delta_{A}^q)^{\frac{1}{q}}\\
\leq&(1-{\rm Tr}\rho_{A}^q+1-{\rm Tr}\delta_{A}^q)^{\frac{1}{q}}\\
\leq&(1-{\rm Tr}\rho_{A}^q)^{\frac{1}{q}}+(1-{\rm Tr}\delta_{A}^q)^{\frac{1}{q}}\\
=&\mathscr{C}_q(|\phi\rangle_{AB})+\mathscr{C}_q(|\varphi\rangle_{AB}).\\
\end{array}
\end{equation}
Here $\rho_A$ and $\delta_A$ are respectively the reduced density matrixes of $|\phi\rangle_{AB}$ and $|\varphi\rangle_{AB}$, the first inequality holds because $C_q(|\phi\rangle_{AB})=1-{\rm Tr}(\rho_A^q)$ satisfies subadditivity \cite{40}, the second inequality can be obtained according to the relation $(a+b)^\beta\leq a^\beta+b^\beta$ for $0\leq a,b\leq1$ and $0<\beta<1$.

The subadditivity of $G_q$-concurrence for arbitrary bipartite mixed states can be verify based on the convexity of $\mathscr{C}_q(\rho_{AB})$ and the relation derived in inequality (\ref{5}). $\hfill\blacksquare$

\section{Analytic formula}\label{III}
Concurrence, as a special measure of $G_q$-concurrence, is of paramount significance in remote entanglement distribution protocols \cite{49}. Therefore, it makes sense to develop an analytic formula to associate $G_q$-concurrence with concurrence.

Let $|\phi\rangle_{AB}$ be a pure state on Hilbert space $\mathcal{H}^2\otimes\mathcal{H}^d$ $(d\geq2)$ (or especially be a two-qubit pure state with Schmidt decomposition $|\phi\rangle_{AB}=\sum_{i=1}^2\sqrt{\lambda_i}|ii\rangle_{AB}$), its $G_q$-concurrence is expressed as
\begin{equation*}
\begin{array}{rl}
\mathscr{C}_q(|\phi\rangle_{AB})=(1-\lambda_1^q-\lambda_2^q)^{\frac{1}{q}},
\end{array}
\end{equation*}
and its concurrence is
\begin{equation*}
\begin{array}{rl}
C(|\phi\rangle_{AB})=2\sqrt{\lambda_1\lambda_2}.
\end{array}
\end{equation*}
Moreover, it is easy to testify that for any $2\otimes d$ pure state there is an analytic function that relates $G_q$-concurrence ($q>1$) to concurrence, namely,
\begin{equation}\label{11}
\begin{array}{rl}
\mathscr{C}_q(|\phi\rangle_{AB})=h_q[C(|\phi\rangle_{AB})],\\
\end{array}
\end{equation}
where $h_q(x)$ is
\begin{equation*}
\begin{array}{rl}
h_q(x)=[1-(\frac{1+\sqrt{1-x^2}}{2})^q-(\frac{1-\sqrt{1-x^2}}{2})^q]^{\frac{1}{q}}\\
\end{array}
\end{equation*}
for $0\leq x\leq1$.

To show that the function $h_q(x)$ is valid for any two-qubit mixed state, we require to analyze under what conditions $h_q(x)$ is a monotonically increasing and convex function on $0\leq x\leq1$.

{\bf Proposition 7}. The function $h_q(x)$ is monotonically increasing with respect to $x$ for $q>1$.

{\bf Proof}. This proposition is true if the first derivative of $h_q(x)$ is nonnegative. Then we derive
\begin{equation*}
\begin{array}{rl}
\frac{dh_q(x)}{dx}=&\frac{1}{2^q}[1-(\frac{1+\sqrt{1-x^2}}{2})^q-(\frac{1-\sqrt{1-x^2}}{2})^q]^{\frac{1}{q}-1}\\
&\times\frac{x[(1+\sqrt{1-x^2})^{q-1}-(1-\sqrt{1-x^2})^{q-1}]}{\sqrt{1-x^2}}.\\
\end{array}
\end{equation*}
Obviously, there is $\frac{dh_q(x)}{dx}>0$ for $0<x<1$ and $q>1$. Therefore, we can say $h_q(x)$ is monotonically increasing for $0\leq x\leq 1$ owing to the fact that $h_q(x)$ is continuous. This makes that $h_q(0)=0$ and $h_q(1)=(1-\frac{1}{2^{q-1}})^{\frac{1}{q}}$ correspond, respectively, to the minimum and maximum of $h_q(x)$ for the given parameter $q$. $\hfill\blacksquare$

{\bf Proposition 8}. The function $h_q(x)$ is convex with respect to $x$ for $1<q\leq2$.

{\bf Proof}. This proposition is valid if the second derivative of $h_q(x)$ is nonnegativity. By derivation, we obtain
\begin{equation*}
\begin{array}{rl}
\frac{d^2h_q(x)}{dx^2}=\frac{1}{2^q}[1-(\frac{1+\sqrt{1-x^2}}{2})^q-(\frac{1-\sqrt{1-x^2}}{2})^q]^{\frac{1}{q}-2}M(x,q),\\
\end{array}
\end{equation*}
where $M(x,q)=\xi_1+\xi_2(\xi_3-\xi_4)$ with
\begin{equation*}
\begin{array}{rl}
&\xi_1=\frac{1-q}{2^q}(\frac{x[(1+\sqrt{1-x^2})^{q-1}-(1-\sqrt{1-x^2})^{q-1}]}{\sqrt{1-x^2}})^2,\\
&\xi_2=1-(\frac{1+\sqrt{1-x^2}}{2})^q-(\frac{1-\sqrt{1-x^2}}{2})^q,\\
&\xi_3=\frac{(1+\sqrt{1-x^2})^{q-2}}{1-x^2}[\frac{1+\sqrt{1-x^2}}{\sqrt{1-x^2}}-x^2(q-1)],\\
&\xi_4=\frac{(1-\sqrt{1-x^2})^{q-2}}{1-x^2}[\frac{1-\sqrt{1-x^2}}{\sqrt{1-x^2}}+x^2(q-1)].\\
\end{array}
\end{equation*}
We observe that judging the sign of $\frac{d^2h_q(x)}{dx^2}$ is actually equivalent to judging the sign of $M(x,q)$ because the term in front of $M(x,q)$ is positive for $0<x<1$.
\begin{figure}[htbp]
\centering
{\includegraphics[width=8cm,height=6cm]{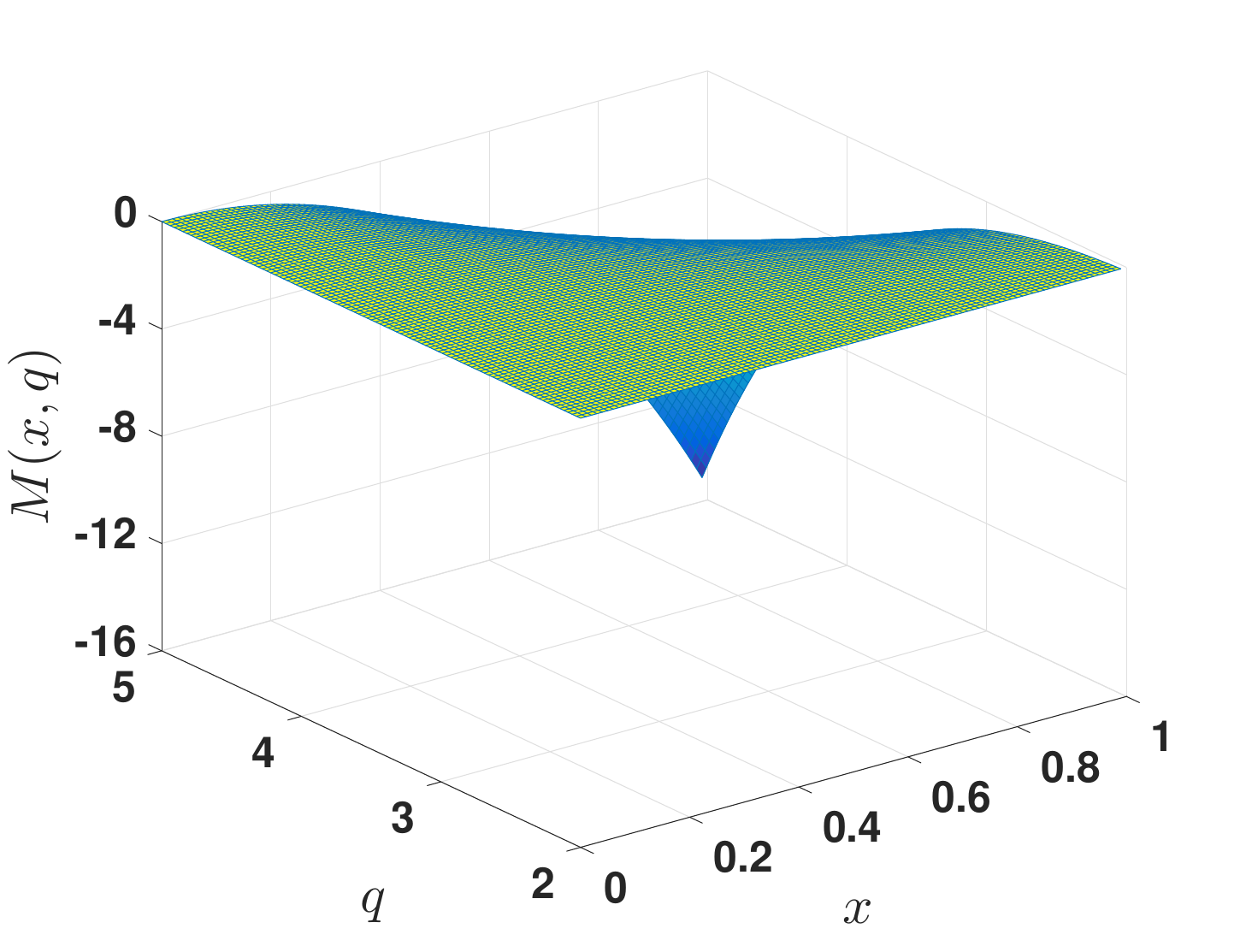}}
\caption{The function $M(x,q)$ is illustrated for $0\leq x\leq1$ and $2<q<5$.} \label{fig 1}
\end{figure}
If the term $\frac{1+\sqrt{1-x^2}}{\sqrt{1-x^2}}-x^2(q-1)$ in $\xi_3$ is non-positive, then there must be $M(x,q)<0$ since $\xi_1<0$, $\xi_2>0$, $\xi_3-\xi_4<0$ for $0<x<1$. By means of the result in Ref. \cite{8}, we deduce directly there is $x\in(0,1)$ such that $\frac{d^2h_q(x)}{dx^2}<0$ for $q\geq5$.  In addition, when $2<q<5$, we can get that $h_q(x)$ is also not a convex function from Fig. \ref{fig 1}. Especially, when $q$ takes 3 and 4, $h_3(x)$ and $h_4(x)$ are respectively
\begin{equation*}
\begin{array}{rl}
h_3(x)=\sqrt[3]{\frac{3}{4}}x^{\frac{2}{3}}~~{\rm and}~h_4(x)=\frac{\sqrt[4]{8x^2-x^4}}{\sqrt[4]{8}}.
\end{array}
\end{equation*}
Obviously, they are concave functions of $x$. Consequently, $h_q(x)$ is not a convex function for $q>2$.

In particular, when $q=2$, $h_2(x)=\frac{x}{\sqrt2}$ is a function that is both convex and concave.

Let us now show that $h_q(x)$ is a convex function on $x\in[0,1]$ for $1<q<2$, that is, prove that the minimum of $M(x,q)$ is nonnegative. It is acknowledged that the minimum value can only be generated at critical points or boundary points since $M(x,q)$ is continuous.

We first discuss whether there are critical points of $M(x,q)$ in the region $R=\{(x,q)|0<x<1,1<q<2\}$. The gradient of $M(x,q)$ is
\begin{equation*}
\begin{array}{rl}
\nabla M(x,q)=\big(\frac{\partial M(x,q)}{\partial x},\frac{\partial M(x,q)}{\partial q}\big),
\end{array}
\end{equation*}
where $\frac{\partial M(x,q)}{\partial x}$ and $\frac{\partial M(x,q)}{\partial q}$ are the first partial derivatives of $M(x,q)$ with respect to $x$ and $q$, respectively.

\begin{figure}[htbp]
\centering
{\includegraphics[width=8cm,height=6cm]{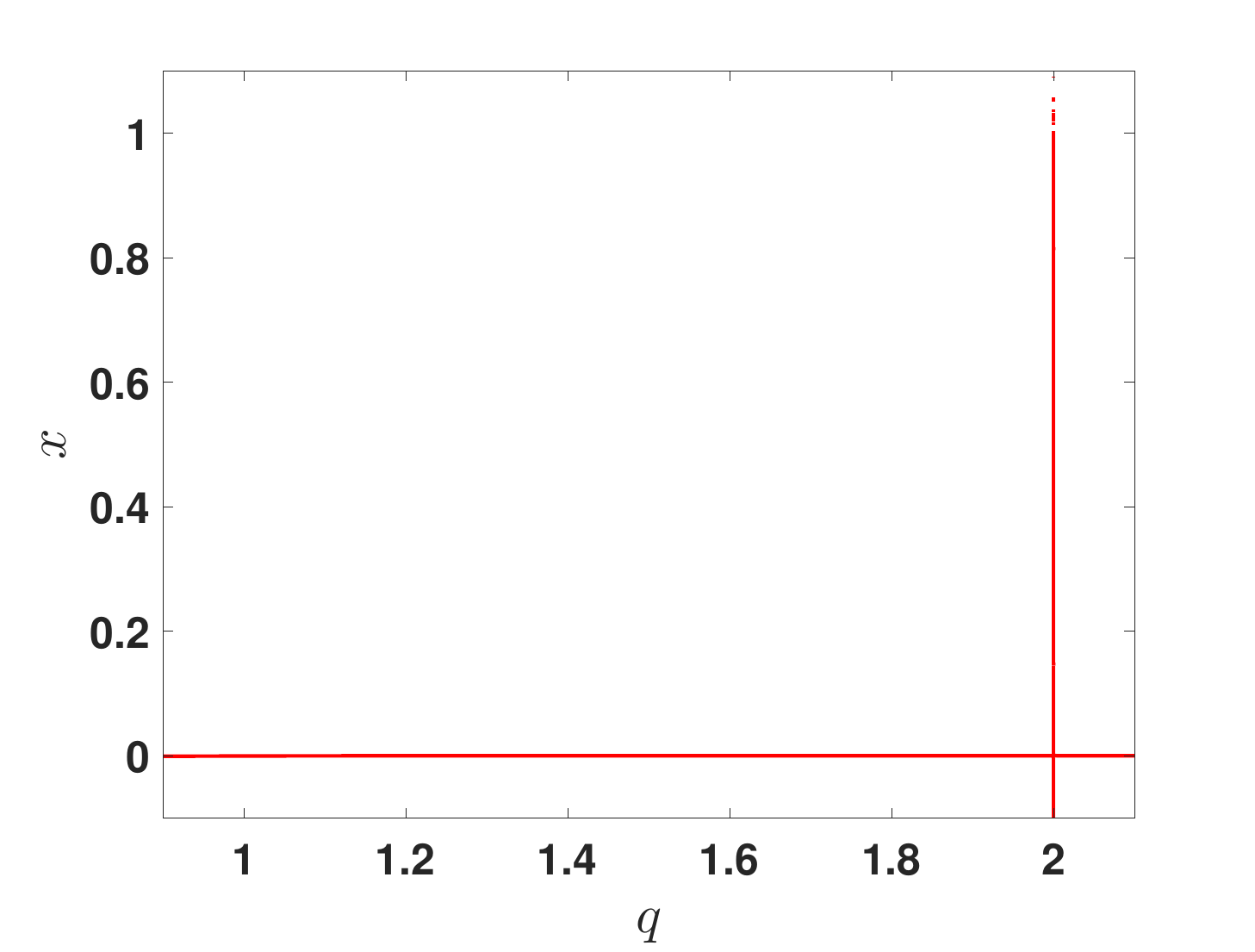}}
\caption{This diagram shows the solution for $\frac{\partial M(x,q)}{\partial x}=0$.}\label{Fig 6}
\end{figure}

\begin{figure}[htbp]
\centering
{\includegraphics[width=8cm,height=6cm]{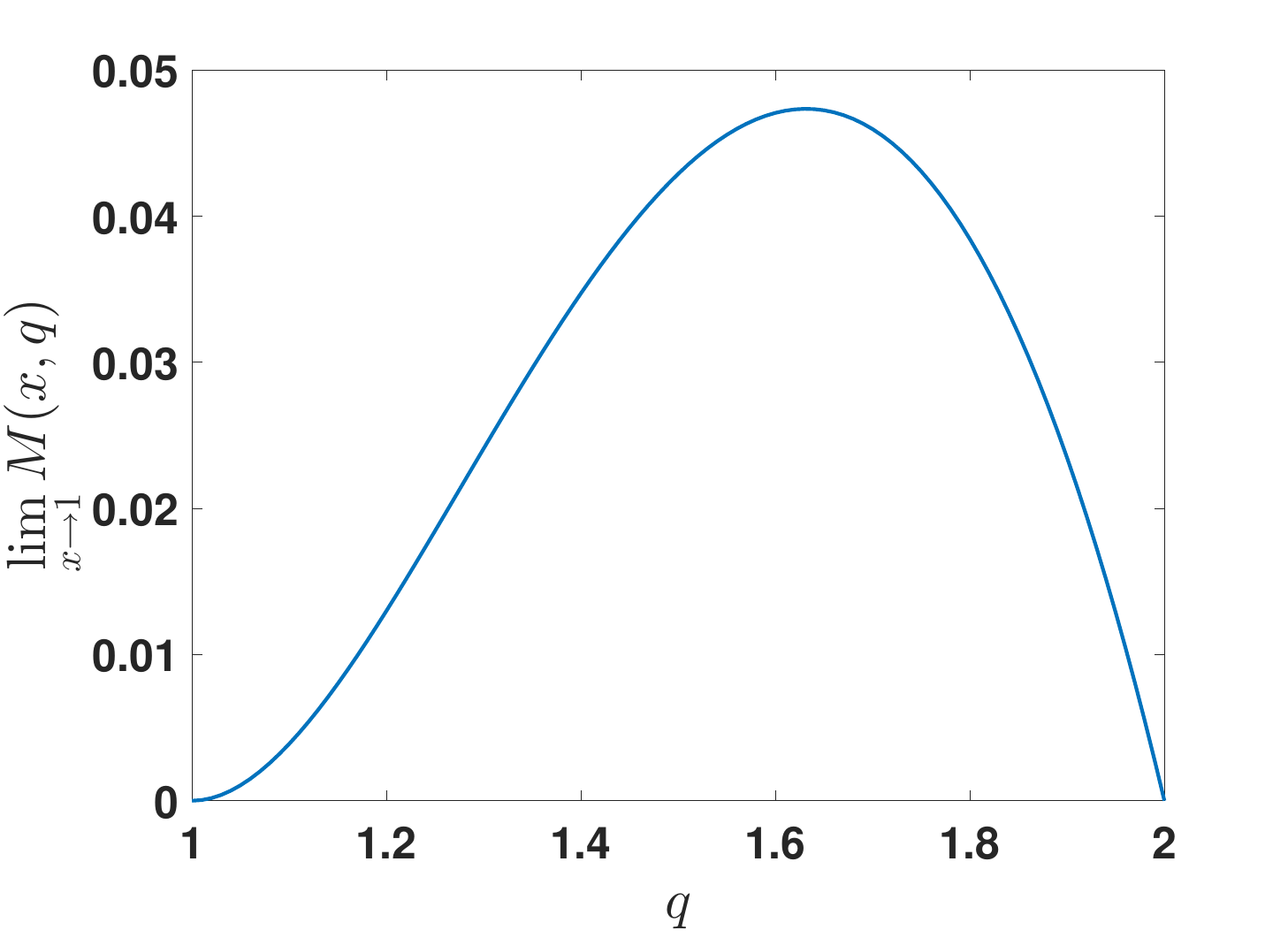}}
\caption{The function $\lim\limits_{x\rightarrow 1}M(x,q)$ is plotted for $1\leq q\leq2$.} \label{fig 2}
\end{figure}

The point $(x_0,q_0)$ is a critical point if $\nabla M(x_0,q_0)=0$. However, $\frac{\partial M(x,q)}{\partial x}=0$ is unsolvable for $0<x<1$ and $1<q<2$, as shown in Fig. \ref{Fig 6}. This suggests that there is no critical point of $M(x,q)$ in the region $R$ and the maximum and minimum values of $M(x,q)$ can be respectively obtained at the boundary points $x=1$ and $x=0$ for given $q$. Through tedious calculations, we get
\begin{equation*}
\begin{array}{rl}
&\lim\limits_{x\rightarrow 0}M(x,q)=0,\\
&\lim\limits_{x\rightarrow 1}M(x,q)=\frac{-12(q^3-3q^2+3q-1)+(2^q-2)(-2q^3+12q^2-16q+6)}{3\times2^q}.
\end{array}
\end{equation*}
The solutions of $\lim\limits_{x\rightarrow 1}M(x,q)=0$ only rise at $q=1$ or 2 for $1\leq q\leq2$, $\lim\limits_{x\rightarrow 1}M(x,q)$ is strictly positive for $1<q<2$, as shown in Fig. \ref{fig 2}. Therefore, one sees $M(x,q)>0$ for $0<x\leq1$ and $1<q<2$.

Based on the above analysis, we can directly obtain $\frac{d^2h_q(x)}{dx^2}>0$ in the region $R$. We proceed to discuss the second derivation $\frac{d^2h_q(x)}{dx^2}$ at the points $x=0$ and $x=1$, and get
\begin{equation*}
\begin{array}{rl}
&\lim\limits_{x\rightarrow 0}\frac{d^2h_q(x)}{dx^2}=+\infty,\\
&\lim\limits_{x\rightarrow 1}\frac{d^2h_q(x)}{dx^2}=\frac{(2^q-2)^{\frac{1-2q}{q}}}{2^{1-q}}\lim\limits_{x\rightarrow 1}M(x,q).\\
\end{array}
\end{equation*}
This makes $\frac{d^2h_q(x)}{dx^2}>0$ for $x\in[0,1]$ and $q\in(1,2)$.

To sum up, the function $h_q(x)$ is convex for $1<q\leq2$ and $0\leq x\leq1$. $\hfill\blacksquare$

Based on propositions 7 and 8, the conclusion can be drawn as follows.

{\bf Theorem 1}. For any two-qubit mixed state $\rho_{AB}$, $G_q$-concurrence has an analytic formula which is a function with regard to concurrence,
\begin{equation}\label{10}
\begin{array}{rl}
\mathscr{C}_q(\rho_{AB})=h_q[C(\rho_{AB})],
\end{array}
\end{equation}
where $1<q\leq2$.

{\bf Proof}. Let $\{p_i,|\phi_i\rangle\}$ be the optimal pure state decomposition of $\mathscr{C}_q(\rho_{AB})$, then one has
\begin{equation*}
\begin{array}{rl}
\mathscr{C}_q(\rho_{AB})&=\sum_ip_i\mathscr{C}_q(|\phi_i\rangle)\\
&=\sum_ip_ih_q[C(|\phi_i\rangle)]\\
&\geq h_q[\sum_ip_iC(|\phi_i\rangle)]\\
&\geq h_q[C(\rho_{AB})],\\
\end{array}
\end{equation*}
where the first inequality is due to the convexity of $h_q(x)$ for $1<q\leq2$ and the second inequality follows from the monotonicity of $h_q(x)$ for $q>1$.

Hill and Wootters \cite{9} pointed out that there is an optimal pure state decomposition $\{p_i, |\phi_i\rangle\}$ for any two-qubit mixed state $\rho_{AB}$ such that the concurrence of each pure state is equal. Based on this assertion, one derives
\begin{equation*}
\begin{array}{rl}
h_q[C(\rho_{AB})]&=h_q[\sum_ip_iC(|\phi_i\rangle)]\\
&=\sum_ip_ih_q[C(|\phi_i\rangle)]\\
&=\sum_ip_i\mathscr{C}_q(|\phi_i\rangle)\\
&\geq\mathscr{C}_q(\rho_{AB}),\\
\end{array}
\end{equation*}
where the inequality holds according to the definition of $\mathscr{C}_q(\rho_{AB})$.

Therefore, the Eq. (\ref{10}) is true for any two-qubit mixed state. $\hfill\blacksquare$

We can straightforwardly arrive at the following result from Theorem 1.

{\bf Corollary 1}. For any mixed state $\rho_{AB}$ in $2\otimes d$ systems, there is
\begin{equation}\label{20}
\begin{array}{rl}
\mathscr{C}_q(\rho_{AB})\geq h_q[C(\rho_{AB})]
\end{array}
\end{equation}
for $1<q\leq2$.

In particular, it is worth noting that formulas (\ref{10}) and (\ref{20}) hold for any bipartite quantum state when $q=2$, not limited to the preconditions given by Theorem 1 and Corollary 1.
\section{Polygamy relation}\label{IV}
It is well-known that the square of concurrence of assistance (CoA) satisfies polygamy relation for $n$-qubit pure state $|\phi\rangle_{A_1A_2\cdots A_n}$, which reads \cite{38}
\begin{equation}\label{19}
\begin{array}{rl}
C^2(|\phi\rangle_{A_1|A_2\cdots A_n})\leq (C^a_{A_1A_2})^2+\cdots+(C^a_{A_1A_n})^2,
\end{array}
\end{equation}
where $C(|\phi\rangle_{A_1|A_2\cdots A_n})$ is the concurrence of $|\phi\rangle_{A_1A_2\cdots A_n}$ under bipartite splitting $A_1|A_2\cdots A_n$, and $C^a_{A_1A_j}$ is the CoA of reduced density operator $\rho_{A_1A_j}$, $j=2,\cdots,n$. Based on inequality (\ref{19}), we will establish the polygamy relation of $n$-qubit state in terms of $G_q$-CoA defined in Eq. (\ref{23}).

To facilitate the proof of polygamy relation, we consider a function of two variables
\begin{equation*}
\begin{array}{rl}
H_q(x,y)=h_q(\sqrt{x^2+y^2})-h_q(x)-h_q(y)\\
\end{array}
\end{equation*}
on the region $R'=\{(x,y)|0\leq x,y,x^2+y^2\leq1\}$ for $1<q\leq2$.

For the case $q=2$, by simple calculation, we have $H_2(x,y)\leq0$.
\begin{figure}[htbp]
\centering
{\includegraphics[width=8cm,height=6cm]{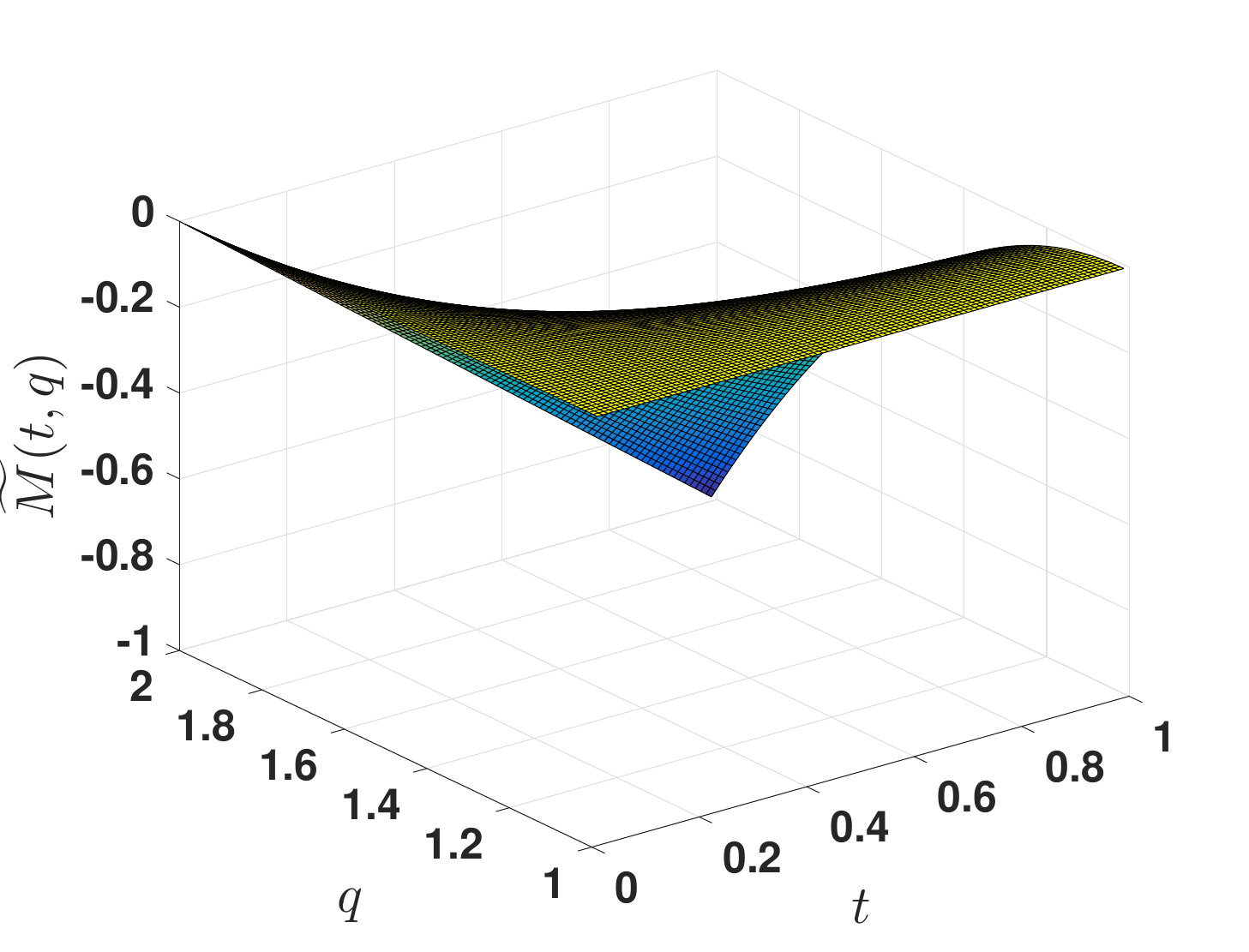}}
\caption{The $\widetilde{M}(t,q)$ is plotted as a function of $t$ and $q$ for $t\in(0,1)$ and $q\in(1,2)$.} \label{fig 7}
\end{figure}

Since $H_q(x,y)$ is continuous on bounded closed set $R'$, it can take maximum and minimum values for given $q$, which occur only at critical or boundary points. First, we determine whether there are critical points in the interior of $R'$ by taking the first partial derivative of $H_q(x,y)$, its gradient is
\begin{equation*}
\begin{array}{rl}
\nabla H_q(x,y)=\big(\frac{\partial H_q(x,y)}{\partial x},\frac{\partial H_q(x,y)}{\partial y}\big),
\end{array}
\end{equation*}
where
\begin{equation*}
\begin{array}{rl}
\frac{\partial H_q(x,y)}{\partial x}=&\frac{x}{2^q}\{[1-(\frac{1+\sqrt{1-x^2-y^2}}{2})^q-(\frac{1-\sqrt{1-x^2-y^2}}{2})^q]^{\frac{1}{q}-1}\\
&\times\frac{(1+\sqrt{1-x^2-y^2})^{q-1}-(1-\sqrt{1-x^2-y^2})^{q-1}}{\sqrt{1-x^2-y^2}}\\
&-[1-(\frac{1+\sqrt{1-x^2}}{2})^q-(\frac{1-\sqrt{1-x^2}}{2})^q]^{\frac{1}{q}-1}\\
&\times\frac{(1+\sqrt{1-x^2})^{q-1}-(1-\sqrt{1-x^2})^{q-1}}{\sqrt{1-x^2}}\},\\
\frac{\partial H_q(x,y)}{\partial y}=&\frac{y}{2^q}\{[1-(\frac{1+\sqrt{1-x^2-y^2}}{2})^q-(\frac{1-\sqrt{1-x^2-y^2}}{2})^q]^{\frac{1}{q}-1}\\
&\times\frac{(1+\sqrt{1-x^2-y^2})^{q-1}-(1-\sqrt{1-x^2-y^2})^{q-1}}{\sqrt{1-x^2-y^2}}\\
&-[1-(\frac{1+\sqrt{1-y^2}}{2})^q-(\frac{1-\sqrt{1-y^2}}{2})^q]^{\frac{1}{q}-1}\\
&\times\frac{(1+\sqrt{1-y^2})^{q-1}-(1-\sqrt{1-y^2})^{q-1}}{\sqrt{1-y^2}}\}.\\
\end{array}
\end{equation*}

Assume that there is $(x_0,y_0)\in\{(x,y)|0<x,y,x^2+y^2<1\}$ such that $\nabla H_q(x_0,y_0)=0$, and we observe that $\nabla H_q(x_0,y_0)=0$ is equivalent to
\begin{equation*}
\begin{array}{rl}
f_q(x_0)=f_q(y_0),
\end{array}
\end{equation*}
where the function $f_q(t)$ is
\begin{equation*}
\begin{array}{rl}
f_q(t)=&[1-(\frac{1+\sqrt{1-t^2}}{2})^q-(\frac{1-\sqrt{1-t^2}}{2})^q]^{\frac{1}{q}-1}\\
&\times\frac{(1+\sqrt{1-t^2})^{q-1}-(1-\sqrt{1-t^2})^{q-1}}{\sqrt{1-t^2}}.\\
\end{array}
\end{equation*}
Then we evaluate the first derivative of $f_q(t)$,
\begin{equation*}
\begin{array}{rl}
\frac{df_q(t)}{dt}=[1-(\frac{1+\sqrt{1-t^2}}{2})^q-(\frac{1-\sqrt{1-t^2}}{2})^q]^{\frac{1}{q}-2}\widetilde{M}(t,q),\\
\end{array}
\end{equation*}\
where
\begin{figure}[htbp]
\centering
{\includegraphics[width=8cm,height=6cm]{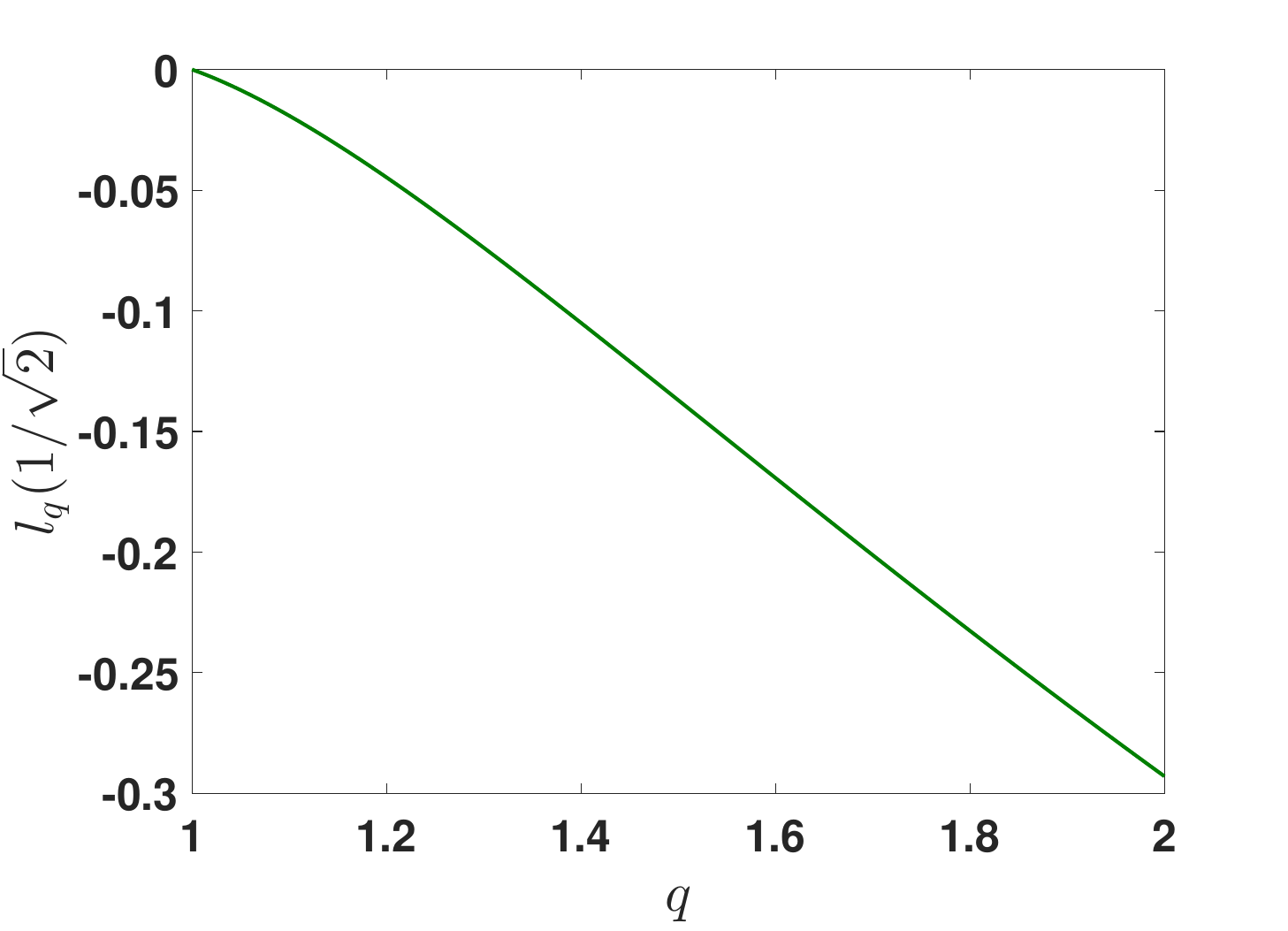}}
\caption{The function $l_q(1/\sqrt{2})$ is non-positive for $1\leq q\leq2$.} \label{fig 3}
\end{figure}
\begin{equation*}
\begin{array}{rl}
\widetilde{M}(t,q)=&\frac{1-q}{2^q}\frac{t[(1+\sqrt{1-t^2})^{q-1}-(1-\sqrt{1-t^2})^{q-1}]^2}{1-t^2}\\
&+[1-(\frac{1+\sqrt{1-t^2}}{2})^q-(\frac{1-\sqrt{1-t^2}}{2})^q]\\
&\times\big[\frac{(1+\sqrt{1-t^2})^{q-2}}{1-t^2}\big(\frac{t(1+\sqrt{1-t^2})}{\sqrt{1-t^2}}-t(q-1)\big)\\
&-\frac{(1-\sqrt{1-t^2})^{q-2}}{1-t^2}\big(\frac{t(1-\sqrt{1-t^2})}{\sqrt{1-t^2}}+t(q-1)\big)\big].\\
\end{array}
\end{equation*}
Combining Fig. \ref{fig 7}, $\lim\limits_{t\rightarrow0}\widetilde{M}(t,q)=0$, and $\lim\limits_{q\rightarrow1}\widetilde{M}(t,q)=0$, we can see $\widetilde{M}(t,q)<0$ for $1<q<2$ and $0<t<1$. This implies $\frac{df_q(t)}{dt}<0$ for $1<q<2$ and $0<t<1$, namely, $f_q(t)$ is a strictly monotonically decreasing function with respect to $t$ for given $q$, so $f_q(x_0)=f_q(y_0)$ means $x_0=y_0$. If $\frac{\partial H_q(x,y)}{\partial x}|_{(x_0,y_0)}=0$ and $x_0=y_0>0$, then $f_q(\sqrt{2}x_0)=f_q(x_0)$, which contradicts to the strict monotonicity of $f_q(t)$. Hence $H_q(x,y)$ has no vanishing gradient in the interior of $R'$.

Next we discuss the boundary values of $H_q(x,y)$ in region $R'$. If $x=0$ or $y=0$, then $H_q(x,y)=0$. If $x^2+y^2=1$, then $H_q(x,y)$ can be reduced to
\begin{equation*}
\begin{array}{rl}
l_q(x)=&H_q(x,\sqrt{1-x^2})\\
=&\frac{1}{2}\{(2^q-2)^{\frac{1}{q}}-[2^q-(1+\sqrt{1-x^2})^q-(1\\
&-\sqrt{1-x^2})^q]^{\frac{1}{q}}-[2^q-(1+x)^q-(1-x)^q]^{\frac{1}{q}}\}.\\
\end{array}
\end{equation*}

For the sake of judging the sign of $l_q(x)$, we calculate its first derivative
\begin{equation*}
\begin{array}{rl}
\frac{dl_q(x)}{dx}=&\frac{1}{2}\{-[2^q-(1+\sqrt{1-x^2})^q-(1-\sqrt{1-x^2})^q]^{\frac{1}{q}-1}\\
&\frac{x[(1+\sqrt{1-x^2})^{q-1}-(1-\sqrt{1-x^2})^{q-1}]}{\sqrt{1-x^2}}-[2^q-(1+x)^q\\
&-(1-x)^q]^{\frac{1}{q}-1}[(1-x)^{q-1}-(1+x)^{q-1}]\},\\
\end{array}
\end{equation*}
where $\frac{dl_q(x)}{dx}=0$ corresponds to $x=\frac{1}{\sqrt{2}}$ on $0<x<1$. Since $l_q(0)=l_q(1)=0$, the sign of function $l_q(x)$ is determined by $l_q(\frac{1}{\sqrt{2}})=\frac{1}{2}\{(2^q-2)^{\frac{1}{q}}-2[2^q-(1+\frac{1}{\sqrt2})^q-(1-\frac{1}{\sqrt2})^q]^{\frac{1}{q}}\}$. We plot $l_q(\frac{1}{\sqrt{2}})$ in the Fig. \ref{fig 3} and obtain that $l_q(\frac{1}{\sqrt{2}})$ is always non-positive for $1<q<2$.

Therefore, we have
\begin{equation}\label{12}
\begin{array}{rl}
h_q(\sqrt{x^2+y^2})\leq h_q(x)+h_q(y)\\
\end{array}
\end{equation}
for $1<q\leq2$ and $(x,y)\in R'$.

Next, by virtue of the inequality (\ref{12}), we consider polygamy relation in multiqubit systems based on $G_q$-CoA.

For the convenience of subsequent proof, we establish a relation between $G_q$-CoA and CoA for two-qubit state $\rho_{AB}$. Let $\{p_i,|\varphi_i\rangle\}$ be the pure state decomposition of $\rho_{AB}$ such that $C^a(\rho_{AB})=\sum_ip_iC(|\varphi_i\rangle)$, then one sees
\begin{equation}\label{13}
\begin{array}{rl}
h_q[C^a(\rho_{AB})]&=h_q[\sum_ip_iC(|\varphi_i\rangle)]\\
&\leq\sum_ip_ih_q[C(|\varphi_i\rangle)]\\
&=\sum_ip_i\mathscr{C}_q(|\varphi_i\rangle)\\
&\leq\mathscr{C}_q^a(\rho_{AB}),\\
\end{array}
\end{equation}
where the first inequality is due to the convexity of $h_q(x)$ for $1<q\leq2$ and the second inequality is assured according to the definition of $\mathscr{C}_q^a(\rho_{AB})$.

{\bf Theorem 2}. For any $n$-qubit quantum state $\rho_{A_1\cdots A_n}$, we have
\begin{equation}\label{17}
\begin{array}{rl}
\mathscr{C}_q^a(\rho_{A_1|A_2\cdots A_n})\leq\mathscr{C}_q^a(\rho_{A_1A_2})+\cdots+\mathscr{C}_q^a(\rho_{A_1A_n})
\end{array}
\end{equation}
for $1<q\leq2$, where $\mathscr{C}_q^a(\rho_{A_1|A_2\cdots A_n})$ denotes the $G_q$-CoA of $\rho_{A_1A_2\cdots A_n}$ in the partition $A_1|A_2\cdots A_n$, $\rho_{A_1A_j}$ is the reduced density matrix with respect to subsystem $A_1A_j$, $j=2,3,\cdots,n$.

{\bf Proof}. On the one hand, if $(C_{A_1A_2}^a)^2+\cdots+(C_{A_1A_n}^a)^2\leq1$ for any multiqubit pure state $|\phi\rangle_{A_1\cdots A_n}$, then one has
\begin{equation}\label{14}
\begin{array}{rl}
&\mathscr{C}_q(|\phi\rangle_{A_1|A_2\cdots A_n})\\
=&h_q[C(|\phi\rangle_{A_1|A_2\cdots A_n})]\\
\leq&h_q\big[\sqrt{(C^a_{A_1A_2})^2+\cdots+(C^a_{A_1A_n})^2}\big]\\
\leq&h_q(C^a_{A_1A_2})+h_q\big[\sqrt{(C^a_{A_1A_3})^2+\cdots+(C^a_{A_1A_n})^2}\big]\\
\leq&\cdots\\
\leq& h_q(C^a_{A_1A_2})+\cdots+h_q(C^a_{A_1A_n})\\
\leq&\mathscr{C}_q^a(\rho_{A_1A_2})+\cdots+\mathscr{C}_q^a(\rho_{A_1A_n}).\\
\end{array}
\end{equation}
Here the first inequality holds according to the formula (\ref{19}) and Proposition 7, the second inequality is true based on formula (\ref{12}), the penultimate inequality is obtained by iterating formula (\ref{12}), and the last inequality can be gotten from the formula (\ref{13}).

On the other hand, if $(C_{A_1A_2}^a)^2+\cdots+(C_{A_1A_n}^a)^2>1$, then there is some $j$ such that $(C_{A_1A_2}^a)^2+\cdots+(C_{A_1A_j}^a)^2\leq1$, whereas $(C_{A_1A_2}^a)^2+\cdots+(C_{A_1A_{j+1}}^a)^2>1$, where $2\leq j\leq n$. Let
\begin{equation*}
\begin{array}{rl}
S=(C_{A_1A_2}^a)^2+\cdots+(C_{A_1A_{j+1}}^a)^2-1,
\end{array}
\end{equation*}
then one reads
\begin{align}\label{15}
&\mathscr{C}_q(|\phi\rangle_{A_1|A_2\cdots A_n}) \notag\\
=&h_q[C(|\phi\rangle_{A_1|A_2\cdots A_n})] \notag\\
\leq& h_q(1) \notag\\
=&h_q\big[\sqrt{(C_{A_1A_2}^a)^2+\cdots+(C_{A_1A_{j+1}}^a)^2-S}\big] \notag\\
\leq& h_q\big[\sqrt{(C_{A_1A_2}^a)^2+\cdots+(C_{A_1A_{j}}^a)^2}\big]\\
&+h_q[\sqrt{(C_{A_1A_{j+1}}^a)^2-S}] \notag\\
\leq& h_q(C_{A_1A_2}^a)+\cdots+h_q(C_{A_1A_j}^a)+h_q(C_{A_1A_{j+1}}^a) \notag\\
\leq&\mathscr{C}^a_q(\rho_{A_1A_2})+\cdots+\mathscr{C}^a_q(\rho_{A_1A_n}).\notag
\end{align}
Here the ideas of proving these inequalities are consistent to that of proving the inequality (\ref{14}) above.

Let $\rho_{A_1A_2\cdots A_n}$ be a multiqubit mixed state and $\{p_i,|\phi_i\rangle_{A_1|A_2\cdots A_n}\}$ be the pure state decomposition such that $\mathscr{C}_q^a(\rho_{A_1|A_2\cdots A_n})=\sum_ip_i\mathscr{C}_q(|\phi_i\rangle_{A_1|A_2\cdots A_n})$. Then one has
\begin{equation}\label{16}
\begin{array}{rl}
&\mathscr{C}_q^a(\rho_{A_1|A_2\cdots A_n})\\
=&\sum\limits_ip_i\mathscr{C}_q(|\phi_i\rangle_{A_1|A_2\cdots A_n})\\
\leq&\sum\limits_ip_i[\mathscr{C}^a_q(\rho^i_{A_1A_2})+\cdots+\mathscr{C}^a_q(\rho^i_{A_1A_n})]\\
=&\sum\limits_ip_i\mathscr{C}^a_q(\rho^i_{A_1A_2})+\cdots+\sum\limits_ip_i\mathscr{C}^a_q(\rho^i_{A_1A_n})\\
\leq&\mathscr{C}^a_q(\rho_{A_1A_2})+\cdots+\mathscr{C}^a_q(\rho_{A_1A_n}).\\
\end{array}
\end{equation}
Here $\rho^i_{A_1A_j}$ is the reduced density matrix of $|\phi_i\rangle_{A_1A_2\cdots A_n}$ with respect to $A_1A_j$, the first inequality can be derived by inequalities (\ref{14}) and (\ref{15}), and the second inequality follows the definition of $G_q$-CoA.

Combining inequalities (\ref{14}), (\ref{15}), and (\ref{16}), we get inequality (\ref{17}) is valid for any $n$-qubit quantum state. $\hfill\blacksquare$

\section{Monogamy relation and entanglement indicators}\label{V}

\subsection{Monogamy relation}
It is well known that the square of concurrence satisfies monogamy relation \cite{11,2}. That is to say, the inequality
\begin{equation}\label{18}
\begin{array}{rl}
C^2(|\phi\rangle_{A_1|A_2\cdots A_n})\geq C^2_{A_1A_2}+\cdots+C^2_{A_1A_n}
\end{array}
\end{equation}
is true for any $n$-qubit pure state $|\phi\rangle_{A_1\cdots A_n}$, where $C_{A_1A_j}$ is the concurrence of reduced density operator $\rho_{A_1A_j}$, $j=2,\cdots,n$. However, the $G_q$-concurrence itself fails for $1<q\leq2$. Consequently, a question arises whether there exists $\alpha$ such that the $\alpha$-th power of $G_q$-concurrence fulfills the monogamy inequality.

Below we elaborate that the square of $G_q$-concurrence obeys monogamy relation for $1<q\leq2$.
\begin{figure}[htbp]
\centering
{\includegraphics[width=8cm,height=6cm]{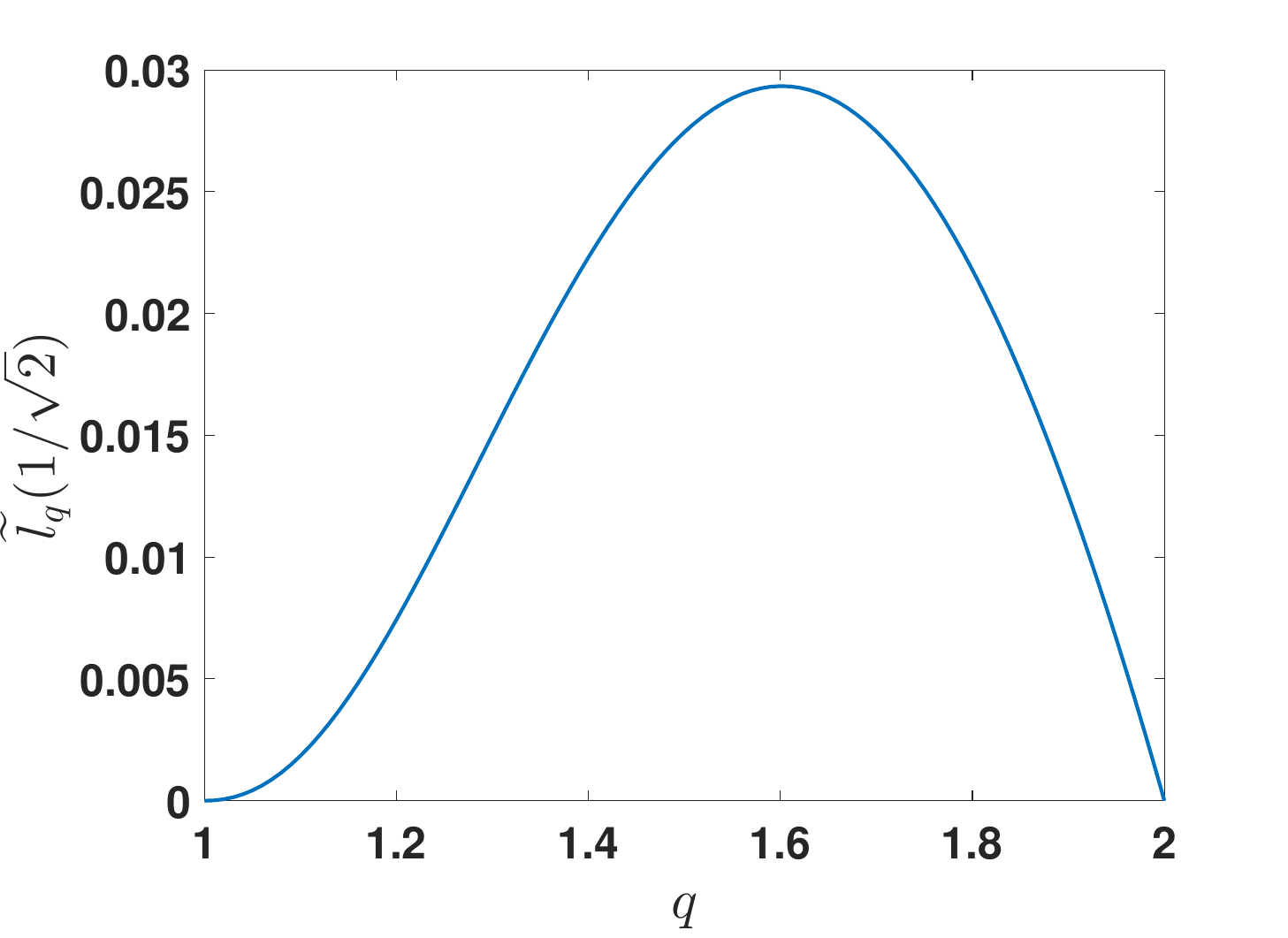}}
\caption{The function $\widetilde{l}_q(1/\sqrt{2})$ is nonnegative for $1<q\leq2$.} \label{fig 4}
\end{figure}

Define
\begin{equation*}
\begin{array}{rl}
\widetilde{H}_q(x,y)=h_q^2(\sqrt{x^2+y^2})-h_q^2(x)-h_q^2(y).\\
\end{array}
\end{equation*}
For the special case $q=2$, it is easy to get $\widetilde{H}_2(x,y)=0$, namely, $h_2^2(\sqrt{x^2+y^2})=h_2^2(x)+h_2^2(y)$. For $1<q<2$, we compute the gradient of $\widetilde{H}_q(x,y)$, denoted $\nabla \widetilde{H}_q(x,y)=\big(\frac{\partial \widetilde{H}_q(x,y)}{\partial x},\frac{\partial \widetilde{H}_q(x,y)}{\partial y}\big)$. Then following similar procedures as in the previous section, we find that $\nabla \widetilde{H}_q(x,y)$ does not disappear in the interior of $R'$ for $1<q<2$. We proceed to consider the boundary values of $\widetilde{H}_q(x,y)$. If $x=0$ or $y=0$, then $\widetilde{H}_q(x,y)=0$. If $x^2+y^2=1$, then
\begin{equation*}
\begin{array}{rl}
\widetilde{l}_q(x)=&\widetilde{H}_q(x,\sqrt{1-x^2})\\
=&\frac{1}{4}\{(2^q-2)^{\frac{2}{q}}-[2^q-(1+\sqrt{1-x^2})^q-(1\\
&-\sqrt{1-x^2})^q]^{\frac{2}{q}}-[2^q-(1+x)^q-(1-x)^q]^{\frac{2}{q}}\}.\\
\end{array}
\end{equation*}
By taking the first derivative of $\widetilde{l}_q(x)$, we find that $\frac{d\widetilde{l}_q(x)}{dx}=0$ corresponds to $x=\frac{1}{\sqrt2}$ for $1<q<2$ and $0<x<1$. Due to $\widetilde{l}_q(0)=\widetilde{l}_q(1)=0$, $\widetilde{l}_q(\frac{1}{\sqrt{2}})$ determines the sign of function $\widetilde{l}_q(x)$. And we observe $\widetilde{l}_q(\frac{1}{\sqrt{2}})\geq0$ from Fig. \ref{fig 4}. Therefore, we have
\begin{equation}\label{6}
\begin{array}{rl}
h_q^2(\sqrt{x^2+y^2})\geq h_q^2(x)+h_q^2(y)\\
\end{array}
\end{equation}
for $1<q\leq2$. It is necessary to mention that inequality (\ref{6}) is strictly greater than if $1<q<2$, $0<x,y<1$, and $x^2+y^2\leq1$.

Before establishing the monogamy relation, we first present a property of the function $h_q^2(x)$.

{\bf Proposition 9}. The function $h_q^2(x)$ is monotonically increasing with respect to $x$ for $0\leq x\leq1$ and $q>1$.

This proposition can be obtained by utilizing the proof of similar approaches with Proposition 7.

{\bf Theorem 3}. For an $n$-qubit quantum state $\rho_{A_1A_2\cdots A_n}$, we have
\begin{equation*}
\begin{array}{rl}
\mathscr{C}_q^2(\rho_{A_1|A_2\cdots A_n})\geq\mathscr{C}_q^2(\rho_{A_1A_2})+\cdots+\mathscr{C}_q^2(\rho_{A_1A_n})
\end{array}
\end{equation*}
for $1<q\leq2$, where $\mathscr{C}_q(\rho_{A_1|A_2\cdots A_n})$ quantifies the entanglement in the partition $A_1|A_2\cdots A_n$.

{\bf Proof}. It is acknowledged that the square of concurrence satisfies monogamy relation and $\mathscr{C}_q(\rho_{A_1|A_2\cdots A_n})=\frac{\sqrt{2}}{{2}}C(\rho_{A_1|A_2\cdots A_n})$ when $q=2$, so $G_2$-concurrence obeys the monogamy relation naturally,
\begin{equation*}
\begin{array}{rl}
\mathscr{C}_2^2(\rho_{A_1|A_2\cdots A_n})\geq\mathscr{C}_2^2(\rho_{A_1A_2})+\cdots+\mathscr{C}_2^2(\rho_{A_1A_n}).\\
\end{array}
\end{equation*}

For any $n$-qubit pure state $|\phi\rangle_{A_1A_2\cdots A_n}$ and $1<q<2$, we derive
\begin{equation}\label{7}
\begin{array}{rl}
\mathscr{C}_q^2(|\phi\rangle_{A_1|A_2\cdots A_n})=&h_q^2[C(|\phi\rangle_{A_1|A_2\cdots A_n})]\\
\geq&h_q^2\big(\sqrt{C_{A_1A_2}^2+\cdots+C_{A_1A_n}^2}\big)\\
\geq&h_q^2(C_{A_1A_2})\\
&+h_q^2\big(\sqrt{C_{A_1A_3}^2+\cdots+C_{A_1A_n}^2}\big)\\
\geq&h_q^2(C_{A_1A_2})+\cdots+h_q^2(C_{A_1A_n})\\
=&\mathscr{C}_q^2(\rho_{A_1A_2})+\cdots+\mathscr{C}_q^2(\rho_{A_1A_n}).\\
\end{array}
\end{equation}
Here the first inequality is because the function $h_q^2(x)$ is monotonically increasing with respect to $x$, the second and third inequalities can be gained by iterating formula (\ref{6}), the last equality follows from Theorem 1.

Given a multiqubit mixed state $\rho_{A_1A_2\cdots A_n}$, we suppose that $\{p_i,|\phi_i\rangle_{A_1|A_2\cdots A_n}\}$ is the optimal pure state decomposition of $\mathscr{C}_q(\rho_{A_1|A_2\cdots A_n})$, that is,  $\mathscr{C}_q(\rho_{A_1|A_2\cdots A_n})=\sum_ip_i\mathscr{C}_q(|\phi_i\rangle_{A_1|A_2\cdots A_n})$. Then we see
\begin{equation*}
\begin{array}{rl}
\mathscr{C}_q^2(\rho_{A_1|A_2\cdots A_n})=&[\sum_ip_i\mathscr{C}_q(|\phi_i\rangle_{A_1|A_2\cdots A_n})]^2\\
=&\{\sum_ip_ih_q[C(|\phi_i\rangle_{A_1|A_2\cdots A_n})]\}^2\\
\geq&\{h_q[\sum_ip_iC(|\phi_i\rangle_{A_1|A_2\cdots A_n})]\}^2\\
\geq&h_q^2[C(\rho_{A_1|A_2\cdots A_n})]\\
\geq&h_q^2(\sqrt{C_{A_1A_2}^2+\cdots+C_{A_1A_n}^2})\\
\geq&\mathscr{C}_q^2(\rho_{A_1A_2})+\cdots+\mathscr{C}_q^2(\rho_{A_1A_n}),\\
\end{array}
\end{equation*}
where the first inequality is assured owing to the convexity of $h_q(x)$ and the monotonicity of the function $y=x^2$ for $x>0$, the second inequality is based on the fact that $h_q^2(x)$ is monotonically increasing and the definition of concurrence, the third inequality is valid because the square of concurrence satisfies monogamy relation for multiqubit quantum states \cite{2}, and the last inequality can be obtained by using the similar procedures with inequality (\ref{7}). $\hfill\blacksquare$

{\bf Corollary 1}. For any $n$-qubit quantum state $\rho_{A_1\cdots A_n}$, the $\alpha$-th ($\alpha\geq2$) power of $G_q$-concurrence fulfills the monogamy relation, i.e.,
\begin{equation*}
\begin{array}{rl}
\mathscr{C}_q^\alpha(\rho_{A_1|A_2\cdots A_n})\geq\mathscr{C}_q^\alpha(\rho_{A_1A_2})+\cdots+\mathscr{C}_q^\alpha(\rho_{A_1A_n}).\\
\end{array}
\end{equation*}

{\bf Proof}. Suppose that $\sum_{i=3}^n\mathscr{C}_q^2(\rho_{A_1A_i})\geq\mathscr{C}_q^2(\rho_{A_1A_2})$, then one derives
\begin{equation*}
\begin{array}{rl}
&\mathscr{C}_q^\alpha(\rho_{A_1|A_2\cdots A_n})\\
&~~\geq[\mathscr{C}_q^2(\rho_{A_1A_2})+\cdots+\mathscr{C}_q^2(\rho_{A_1A_n})]^{\frac{\alpha}{2}}\\
&~~=\Big(\sum_{i=3}^n\mathscr{C}_q^2(\rho_{A_1A_i})\Big)^{\frac{\alpha}{2}}\Big(1+\frac{\mathscr{C}_q^2(\rho_{A_1A_2})}{\sum_{i=3}^n\mathscr{C}_q^2(\rho_{A_1A_i})}\Big)^{\frac{\alpha}{2}}\\
&~~\geq\Big(\sum_{i=3}^n\mathscr{C}_q^2(\rho_{A_1A_i})\Big)^{\frac{\alpha}{2}}\Big[1+\Big(\frac{\mathscr{C}_q^2(\rho_{A_1A_2})}{\sum_{i=3}^n\mathscr{C}_q^2(\rho_{A_1A_i})}\Big)^{\frac{\alpha}{2}}\Big]\\
&~~=\mathscr{C}_q^\alpha(\rho_{A_1A_2})+\Big(\sum_{i=3}^n\mathscr{C}_q^2(\rho_{A_1A_i})\Big)^{\frac{\alpha}{2}}\\
&~~\geq\mathscr{C}_q^\alpha(\rho_{A_1A_2})+\cdots+\mathscr{C}_q^\alpha(\rho_{A_1A_n}).\\
\end{array}
\end{equation*}
Here the first inequality holds because $\mathscr{C}_q^2(\rho_{A_1|A_2\cdots A_n})$ obeys monogamy relation and $y=x^{\frac{\alpha}{2}}$ is a monotonically increasing function with respect to $x$ for $0\leq x\leq1$ and $\alpha\geq2$, the second inequality is according to the inequality $(1+x)^{\frac{\alpha}{2}}\geq1+x^{\frac{\alpha}{2}}$, and the last inequality is valid since the relation $(\sum_ix_i^2)^{\frac{\alpha}{2}}\geq\sum_ix_i^\alpha$ holds for $0\leq x_i\leq1$ and $\alpha\geq2$. $\hfill\blacksquare$

\subsection{Entanglement indicators}
On account of the relation presented in Theorem 3, we give a set of multipartite entanglement indicators,
\begin{equation}\label{21}
\begin{array}{rl}
\tau_q(\rho)=\min\limits_{\{p_l,|\phi_l\rangle\}}\sum\limits_lp_l\tau_q(|\phi_l\rangle_{A_1|A_2\cdots A_n}),\\
\end{array}
\end{equation}
where the minimum is taken over all feasible ensemble decompositions, $\tau_q(|\phi_l\rangle_{A_1|A_2\cdots A_n})=\mathscr{C}^2_q(|\phi_l\rangle_{A_1|A_2\cdots A_n})-\sum_{j=2}^n\mathscr{C}^2_q(\rho^l_{A_1A_j})$, and  $1<q<2$. For the tripartite entanglement indicator $\tau_q(\rho_{ABC})$, the following result can be obtained.

{\bf Theorem 4}. For any three-qubit quantum state $\rho_{ABC}$, the tripartite entanglement indicator $\tau_q(\rho_{ABC})$ is zero for $1<q<2$ iff $\rho_{ABC}$ is biseparable, namely, $\rho_{ABC}=\sum_ip_i\rho_{AB}^i\otimes\rho_{C}^i+\sum_iq_i\rho_{AC}^i\otimes\rho_{B}^i+\sum_ir_i\rho_{A}^i\otimes\rho_{BC}^i$.

{\bf Proof}. If a three-qubit pure state $|\phi\rangle_{ABC}$ is biseparable, then its forms might be
\begin{equation*}
\begin{array}{rl}
&|\phi\rangle_{ABC}=|\phi\rangle_{AB}\otimes|\phi\rangle_{C},\\
&|\phi\rangle_{ABC}=|\phi\rangle_{AC}\otimes|\phi\rangle_{B},\\
&|\phi\rangle_{ABC}=|\phi\rangle_{A}\otimes|\phi\rangle_{BC},\\
&|\phi\rangle_{ABC}=|\phi\rangle_{A}\otimes|\phi\rangle_{B}\otimes|\phi\rangle_{C}.\\
\end{array}
\end{equation*}
We have $\tau_q(|\phi\rangle_{A|BC})=0$ for these states.

Next, the sufficiency is proven. We will illustrate the fact that there is at most one nonzero two-qubit concurrence for three-qubit pure state if $\tau_q(|\phi\rangle_{A|BC})=0$.

If $\mathscr{C}_q(\rho_{AB})>0$ and $\mathscr{C}_q(\rho_{AC})>0$, then we derive
\begin{equation*}
\begin{array}{rl}
&\mathscr{C}^2_q(|\phi\rangle_{A|BC})-\mathscr{C}^2_q(\rho_{AB})-\mathscr{C}^2_q(\rho_{AC})\\
=&h_q^2[C(|\phi\rangle_{A|BC})]-h_q^2[C(\rho_{AB})]-h_q^2[C(\rho_{AC})]\\
\geq&h_q^2[\sqrt{C^2(\rho_{AB})+C^2(\rho_{AC})}]-h_q^2[C(\rho_{AB})]\\
&-h_q^2[C(\rho_{AC})]\\
>&0,\\
\end{array}
\end{equation*}
which is contradictory to the precondition $\tau_q(|\phi\rangle_{A|BC})=0$, where the second inequality is attained due to $h_q^2(\sqrt{x^2+y^2})>h_q^2(x)+h_q^2(y)$ for $1<q<2$, $0<x,y<1$, and $x^2+y^2\leq1$.

If $\mathscr{C}_q(\rho_{AB})=\mathscr{C}_q(\rho_{AC})=\mathscr{C}_q(|\phi\rangle_{A|BC})=0$, then the state may be in the forms $|\phi\rangle_{ABC}=|\phi\rangle_{A}\otimes|\phi\rangle_{BC}$ or
$|\phi\rangle_{ABC}=|\phi\rangle_{A}\otimes|\phi\rangle_{B}\otimes|\phi\rangle_{C}$.

According to the strict concavity of Tsallis entropy and $y=x^\gamma$ $(0<\gamma<1)$, it is not difficult to show that $G_q(\rho)$ is also a strictly concave function.

If there is only one nonzero two-qubit concurrence, $\mathscr{C}_q(\rho_{AB})>0$ or $\mathscr{C}_q(\rho_{AC})>0$, based on the strict concavity of $G_q(\rho)$ and by means of similar procedures to that in Ref. \cite{13}, we get that the corresponding forms of states are $|\phi\rangle_{ABC}=|\phi\rangle_{AB}\otimes|\phi\rangle_{C}$ and $|\phi\rangle_{ABC}=|\phi\rangle_{AC}\otimes|\phi\rangle_{B}$, respectively.

Based on the above discussion, we can derive easily $\tau_q(\rho_{ABC})=0$ for any three-qubit mixed state $\rho_{ABC}$ iff $\rho_{ABC}$ can be expressed in the form $\rho_{ABC}=\sum_ip_i\rho_{AB}^i\otimes\rho_{C}^i+\sum_iq_i\rho_{AC}^i\otimes\rho_{B}^i+\sum_ir_i\rho_{A}^i\otimes\rho_{BC}^i$. $\hfill\blacksquare$

In fact, the inequality
\begin{equation*}
\begin{array}{rl}
\mathscr{C}^2_q(\rho_{A_i|\overline{A_i}})\geq\sum_{j\neq i}\mathscr{C}^2_q(\rho_{A_iA_j})
\end{array}
\end{equation*}
is true for any $n$-qubit quantum state. As a consequence, we construct a family of indicators
\begin{equation}\label{22}
\begin{array}{rl}
\tau_q^i(\rho)=\min\limits_{\{p_l,|\phi_l\rangle\}}\sum\limits_lp_l\tau_q^i(|\phi_l\rangle_{A_i|\overline{A_i}}),\\
\end{array}
\end{equation}
where the minimum runs over all feasible pure decompositions, $\tau_q^i(|\phi_l\rangle_{A_i|\overline{A_i}})=\mathscr{C}^2_q(|\phi_l\rangle_{A_i|\overline{A_i}})-\sum\limits_{j\neq i}\mathscr{C}^2_q(\rho^l_{A_iA_j})$, $i,j\in\{1,2,\cdots,n\}$. The formula (\ref{21}) is a special case of $\tau_q^i(\rho)$ corresponding to $i=1$.

{\bf Theorem 5}. For any $n$-qubit quantum state $\rho$, the multipartite entanglement indicator $\tau_q^i(\rho)$ is zero for $1<q<2$ iff the quantum sate can be written in the form $\rho=\sum\limits_kp_k\rho_{A_i}^k\otimes\rho_{\overline{A_i}}^k+\sum\limits_{j\neq i}\sum\limits_kq_k^j\rho_{A_iA_j}^k\otimes\rho_{\overline{A_iA_j}}^k$, where $i,j\in\{1,2,\cdots,n\}$.

Using analogous procedures to lemmas b and c in supplementary material of Ref. \cite{13}, we get above conclusion. Such a series of entanglement indicators, $\tau_q^1(\rho)$ to $\tau_q^n(\rho)$, allow us to better understand the entanglement distribution of states and must be nonzero for any genuinely multiqubit entangled state. Therefore, these indicators we construct can compensate for the deficiency that the tangle fails to detect the entanglement of $n$-qubit $W$ state, as the following example shows.

There exists a right-neighborhood of one $(1,1+\delta_1)$ and a left-neighborhood of two $(2-\delta_2,2)$ such that $\tau_q(\rho)$ is strictly greater than zero but very close to zero. For clarity, we discuss in the interval $[1+\delta_1, 2-\delta_2]$.

{\bf Example 1}. For $n$-qubit $W$ state $|W_n\rangle_{A_1\cdots A_n}=\frac{|10\cdots0\rangle+|01\cdots0\rangle+\cdots+|00\cdots1\rangle}{\sqrt{n}}$,
\begin{figure}[htbp]
\centering
{\includegraphics[width=8cm,height=6cm]{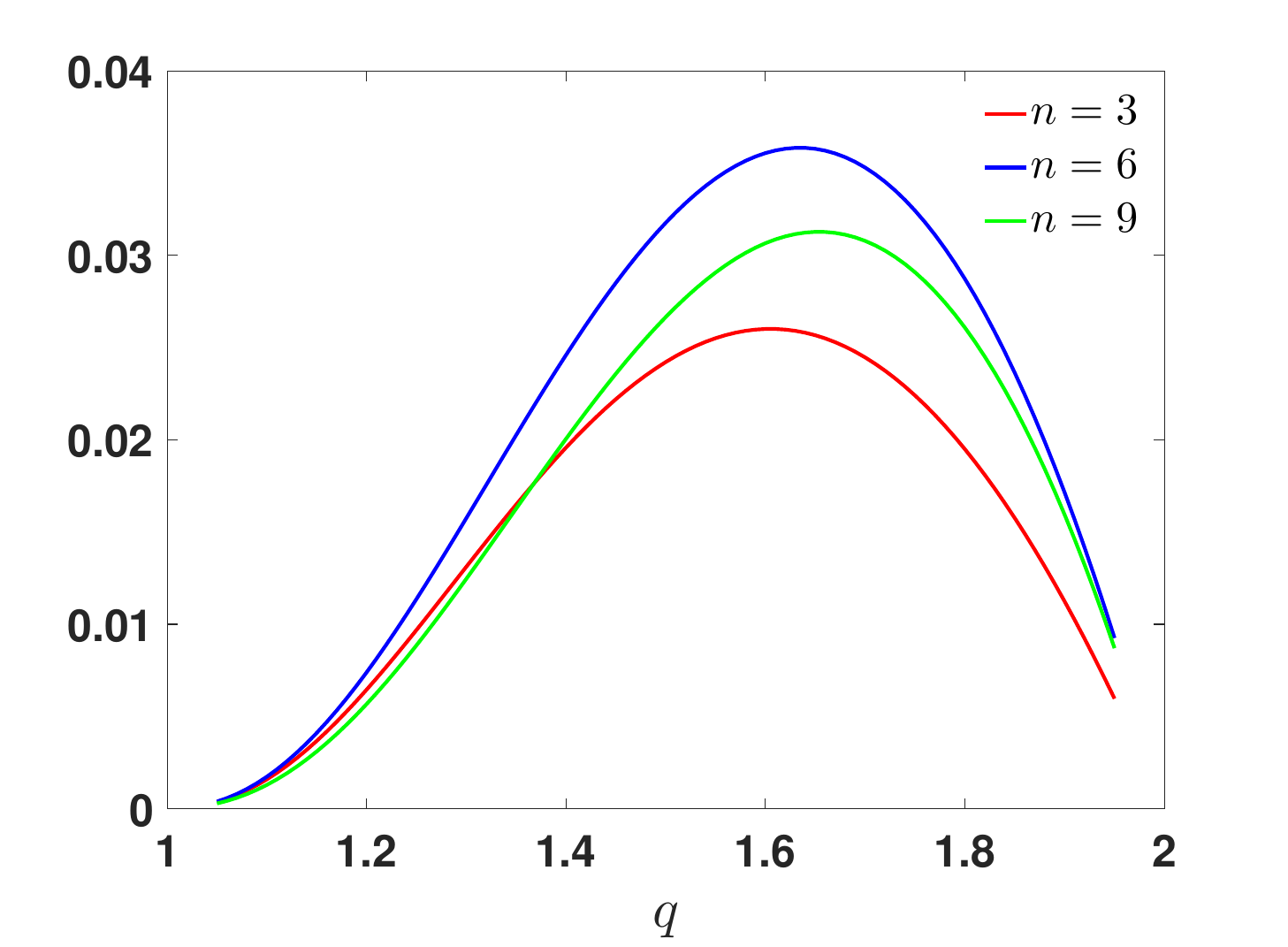}}
\caption{The red line, blue line, and green line correspond to $\tau_q(|W_n\rangle)$ with $n=3,6,9$, where $q\in[1.05,1.95]$.} \label{fig 5}
\end{figure}
its concurrence between subsystems $A_1$ and $A_2\cdots A_n$ is $C(|W_n\rangle_{A_1|A_2\cdots A_n})=\frac{2\sqrt{n-1}}{n}$, $C(\rho_{A_1A_j})=\frac{2}{n}$, $j=2,\cdots,n$, then we have
\begin{equation*}
\begin{array}{rl}
\tau_q(|W_n\rangle)=&\big[1-\big(\frac{1+\frac{n-2}{n}}{2}\big)^q-\big(\frac{1-\frac{n-2}{n}}{2}\big)^q\big]^{\frac{2}{q}}-(n-1)\\
&\big[1-\big(\frac{1+\frac{\sqrt{n^2-4}}{n}}{2}\big)^q-\big(\frac{1-\frac{\sqrt{n^2-4}}{n}}{2}\big)^q\big]^{\frac{2}{q}}.\\
\end{array}
\end{equation*}
Taking $n=3,6,9$ and plotting them in Fig. \ref{fig 5}, we observe $\tau_q(|W_n\rangle)$ is greater than zero obviously for $q\in[1.05,1.95]$.

\section{Conclusion}\label{VI}
In this paper, we propose a type of one-parameter entanglement quantifiers, $G_q$-concurrence ($q>1$), and demonstrate rigorously that they fulfill all axiomatic requirements of an entanglement measure. In addition, an analytic formula between $G_q$-concurrence and concurrence is established for $1<q\leq2$ in two-qubit systems. Furthermore, on account of this analytic formula, we provide the polygamy inequality based on $G_q$-concurrence of assistance in multiqubit systems. Unfortunately, $G_q$-concurrence ($1<q\leq2$) itself does not satisfy the monogamy inequality. A mathematical characterization of monogamy relation in terms of the square of $G_q$-concurrence, however, is presented in multiqubit systems. By virtue of the monogamy inequality, we construct a set of entanglement indicators, which can work well even when the tangle loses its efficacy. Moreover, a straightforward conclusion that the $\alpha$-th $(\alpha\geq2)$ power of $G_q$-concurrence also satisfies monogamy inequality can be reached. The calculation of the multipartite entanglement measures is an NP problem, and these monogamy and polygamy inequalities may provide an alternative perspective on the estimation of entanglement of multiqubit quantum states.

\section*{ACKNOWLEDGMENTS}
This work was supported by the National Natural Science Foundation of China under Grants No. 12071110 and No. 62271189, and the Hebei Central Guidance on Local Science and Technology Development Foundation of China under Grant No. 236Z7604G.


\end{document}